# RunnerDNA: Interpretable indicators and model to characterize human activity pattern and individual difference

Authors: Yao Yao[a,c], Zhuolun Wang[a], Peng Luo[b,*], Hanyu Yin[a], Ziqi Liu[a], Jiaqi Zhang[a], Nengjing Guo[d], Qingfeng Guan[a],

[a] *School of Geography and Information Engineering, China University of Geoscience, Wuhan, Hubei 430078, China;* [b] *Chair of Cartography, Technical University of Munich, 80333 Munich, Germany;* [c] *Alibaba Group, Hangzhou, Zhejiang 311121, China;* [d] *School of Remote Sensing and Information Engineering, Wuhan University, Wuhan, Hubei 430072, China*

[*] Corresponding author

**Corresponding author:**

*Mr. Peng Luo

Chair of Cartography, Technical University of Munich, 80333 Munich, Germany.

Email: peng.luo@tum.de

**Contact information of the co-authors:**

Prof. Yao Yao: yaoy@cug.edu.cn

Mr. Zhuolun Wang: 20171003972@cug.edu.cn

Mr. Hanyu Yin: 20171001660@cug.edu.cn

Mr. Ziqi Liu: liuziqi@cug.edu.cn

Ms. Jiaqi Zhang: 20151001823@cug.edu.cn

Ms. Nengjing Guo: nengjing_guo@whu.edu.cn

Prof. Qingfeng Guan: guanqf@cug.edu.cn

# RunnerDNA: Interpretable indicators and model to characterize human activity pattern and individual difference

**Abstract:** Human activity analysis based on sensor data plays a significant role in behavior sensing, human-machine interaction, health care, and so on. The current research focused on recognizing human activity and posture at the activity pattern level, neglecting the effective fusion of multi-sensor data and assessing different movement styles at the individual level, thus introducing the challenge to distinguish individuals in the same movement. In addition, key features of human activity and individual style are not characterized due to the lack of interpretability of current used models. In this study, the concept of RunnerDNA, consisting of five interpretable indicators, balance, stride, steering, stability, and amplitude, was proposed to describe human activity at the individual level. We collected smartphone multi-sensor data from 33 volunteers who engaged in physical activities such as walking, running, and bicycling and calculated the data into five indicators of RunnerDNA. The indicators were then used to build random forest models and recognize movement activities and the identity of users. The results show that the proposed model has high accuracy in identifying activities (accuracy of 0.679) and is also effective in predicting the identity of running users (with an accuracy of 0.511). Furthermore, the accuracy of the human activity recognition model has a significant improvement (reach to 0.830) by combing RunnerDNA and two motion feature indicators, velocity, and acceleration. Results demonstrate that RunnerDNA is an effective way to describe an individual's physical activity and helps us understand individual differences in sports style, and the significant differences in balance and amplitude between men and women were found. This research provides theoretical support for studying human behavior perception and is conducive to further understanding of human sports behavior and sports style.

**Key words:** RunnerDNA, Human activity recognition, Sport style, Random Forest

**Highlights:**

1. A sport activity recognition model is proposed.

2. Describe human activity and movement style at the individual level by RunnerDNA.

3. Identify movement users and distinguish between different users effectively.

4. Quantitative analysis of differences between male and female users in activity.

# 1. Introduction

Human activity recognition based on sensors draws a lot of attention as sensor technology develops (Zhuang *et al.* 2019). Human activities and style can be revealed by analyzing sensor data, which are extensively used in the fields of mobile health (Kay *et al.* 2011, Su *et al.* 2014), health monitoring (González *et al.* 2015), and the human-machine interaction(Paudyal *et al.* 2016). By quantity analysis of human movement, human behavior and individual movement characteristics effectively and intuitively can be measured and understood.

There are several challenges in human activity recognition, including the method for data integration and collection (Lara and Labrador 2012), the accuracy of recognition algorithm (Hassan *et al.* 2018), and the difficulty of recognizing human activity at the individual scale (Kwan 2004). To better recognize human activity using sensor data, researchers have proposed extensive approaches (Nweke *et al.* 2018, Ramasamy Ramamurthy and Roy 2018).

In terms of data collection of human activity recognition research, some researchers conducted the study on sports activities using a 3-axis accelerometer (Rasekh *et al.* 2014) and orientation sensor

(Martinez-Hernandez *et al.* 2017). Only single-source data are collected in these works, which is not conducive to analyzing the human's comprehensive movement styles. The development of multi-source sensor fusion technology offers the opportunity to collect human activity data from multi-sensor wearable devices. For example, Yuan *et al.* (2017) proposed a model to recognize six sports activities like running, walking, and sitting, using the data from the wearable device. In addition, the GPS sensor in the mobile phone, as a widely used sensor, has been proven the effectiveness for human movement pattern recognition (Kaghyan and Sarukhanyan 2013, Zhang and Poslad 2014). Han *et al.* (2016) obtained human activity data from the smart insole and identified the sports activities of users, including walking, running, and climbing stairs. In spite of the fact that those studies increase the recognition accuracy of sports activities, there are still some problems faced. For instance, the user experience need be improved due to the poor convenience of multi-sensor devices and the complicated data collection process(Al Kalaa and Refai 2015). Moreover, human activity data are collected in an unnatural state, leading to the results deviate from the natural state. Nowadays, the emergence of smartphones with multiple sensors makes up for the lack of wearable sensors. Since smartphones are easy to carry and have become an irreplaceable part of our real life, it has the promising ability to collect human activity data in a natural condition consistently (Dutta *et al.* 2018).

The human activity recognition algorithm should also be improved to obtain a higher recognition accuracy. There have been proposed extensive algorithms for activity recognition (Anguita *et al.* 2012, Shoaib *et al.* 2016), and the deep learning and machine learning method are most commonly used. For example, Wan *et al.* (2020) recognized human activity using a deep learning method based on convolutional neural networks (CNN). The model got a high accuracy on the two datasets, which are 93.31 % and 91.66%, respectively. Deep learning methods are difficult to be explained compared to traditional machine learning methods and are normally called black box model (Yosinski *et al.* 2015). Therefore, it is difficult to perform interpretable analysis of the study results when using deep learning algorithms (Doshi-Velez and Kim 2017). To better explain the model and understand human movement patterns, some high interpretable models are used, include decision tree (DT), naive bayes classifier (NB), support vector machine (SVM), k-Nearest neighbor (KNN), and random forest (RF) for human activity recognition. Previous studies have compared the recognition accuracy of different ML algorithms especially interpretable models to human sports activity. To name a few, Uddin and Uddiny (2015) assessed the performance of KNN, SVM, and improved RF at different feature sets of human sports activity recognition and found improved RF has the highest accuracy. Guinness (2015) used GPS data and accelerator data from smartphone sensors to recognize sports activities, including walking, running, and driving, and compared the accuracy of different algorithms. The result showed that RF has the best performance, with an accuracy of 96.5%. However, although those machine learning algorithms have been widely used to recognize human activities, the individual styles in the same human activity are still under-explored.

While the topic of human activity recognition using multi-sensor data is not new, most researchers focused on the analysis at the activity pattern scale (Hassan *et al.* 2018, Kwapisz *et al.* 2011, Nweke *et al.* 2018), which failed to conduct the sport style analysis at the individual scale. Given that, Fugiglando *et al.* (2017) proposed the concept of DirverDNA, consider individual driving behavior as an overall result of four easily measurable characteristics: braking, steering, speed, and energy efficiency. Their results of the experiment showed that the DriverDNA could better distinguish the drive styles of different drivers. Similar to driving style, people's sports styles also have obvious individual differences (Okamoto *et al.* 2014, Smith *et al.* 2019). However, DriverDNA can only identify driving behaviors, which can't apply to recognizing sports activities, such as running and biking.

In general, the data acquisition process is normally complicated in current studies, and they ignore the effective integration of multi-sensor data. Besides, there is a lack of studies on the finer analysis of movement behavior and movement characteristics at the individual scale, making it impossible to use movement data to identify users and distinguish between different users. To address those problems, we proposed the RunnerDNA based on smartphone multi-sensor data. Using five interpretable indicators of RunnerDNA, random forest models were built to recognize sports activities and the user's identity and analyze sport styles. Based on this, we coupled RunnerDNA and GPS data to further investigate the factors that affect the accuracy of human movement patterns and user identity identification models. The experimental results show that RunnerDNA can effectively identify movement patterns in typical human activities (running, walking, biking, and e-bike riding) and can also profile individual-scale sports styles, as well as identify movement identities, that is, with the activity data, we can estimate the producer of the data. To the best of our knowledge, RunnerDNA presented in this study is the first concept that uses multi-sensor data to describe the human activity, estimate the user's identity and analyze sports style at the individual scale.

## 2. Method

The flow chart of this study is shown in Figure 1, which can be summarized into the following steps: 1) Developing a smartphone APP and using it to collect different sports activity data from volunteers, then aggregated and calculated those data into five indicators of RunnerDNA; 2) Using RunnerDNA to build Random forest models for recognition of user' identity and sports activity; 3) quantitatively analyze the user's sport style according to the results of models; 4) Based on the above analysis, the RunnerDNA was combined with GPS data to analysis the factors to influence the identification accuracy to human movement pattern and user's identity.

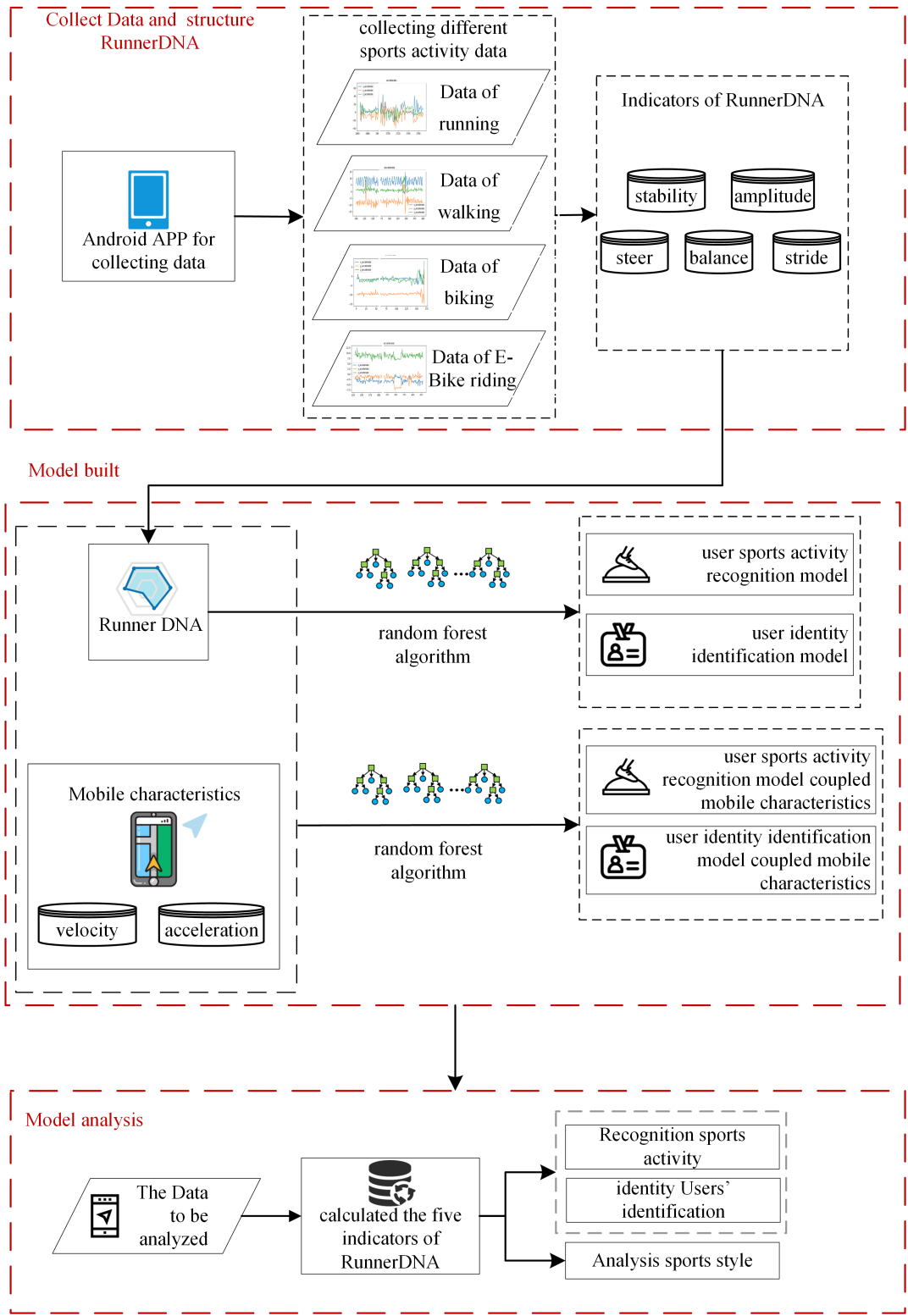

Figure 1 Workflow of sports activity recognition and identification using RunnerDNA

## 2.1 Data acquisition

We developed an Android application for human activity data collection on smartphones. Data generated by six smartphone sensors at 1 HZ were obtained when volunteers were in motion. The basic information of sensors is in Table 1. To exclude as much as the influence of other factors on the results during the experiment, the mobile phones of all volunteers were turned upside down and placed on the front side of the same position of the right leg. In this way, when volunteers are moving forward, the X-axis of phone sensors points to the front left; the Y-axis points perpendicular to the ground.

Table 1 The description of mobile sensors used in this study

| Sensor | Basis functions | Unit |
| --- | --- | --- |
| Accelerometer | Measure three-axis acceleration (x, y, z) including gravity | $m/s^2$ |
| Linear acceleration sensor | Measure three-axis acceleration (x, y, z) without gravity | $m/s^2$ |
| Gravity sensor | Measure the three-axis gravitational acceleration applied to the device (x, y, z) | $m/s^2$ |
| Magnetic sensor | Measure the intensity and direction of the three-axis ambient magnetic field (x, y, z) | $mT$ |
| Orientation sensor | Calculate the angle data of the three-axis direction sensor (x, y, z) | ° |
| Gyroscope | Measure the angular velocity of three axes in space (x, y, z) | $rad/s$ |

Since body shapes like height and weight can significantly impact the data collection, to minimize those impact, we set the volunteer selection criteria as follows: all volunteers are from 18 to 22 years old, men's height range is 167cm~180cm, and weight range is 53kg~76kg; women's height range is 155cm~167cm, the weight range is 45kg~67kg. Volunteers all ride the same shared bikes from Hellobike, and the same shared electric bikes from KVCOOGO. Hellobike and KVCOOGO are two popular bike-sharing companies in China, which offer standard bikes and electric bikes, respectively. Regarding the speed of the different movement patterns, to ensure that the data acquired reflects the natural state of human movement as far as possible, for every kind of sports, we ask volunteers to move at their own normal speed, without deliberate speed control. We also controlled the movement distance of volunteers in the experiment, specifying that each cycling distance should be at least 2km, and each walking and running distance should be at least 800m. Thirty-three volunteers participate in our study. Among them, 20 people are volunteers to running, and others volunteer for other sports activities. The data have 268 records, including 32 records of biking, 55 records of E-Bikes riding, 45 records of walking, and 139 records of running. Each record contains several physical characteristics of volunteers, like sex, height, and weight. It should be mentioned that all data collection processes in this study were informed to the volunteers and the volunteers agreed the data be provided to this study for in-depth analysis. One example record in biking activity is as shown in Table 2.

Table 2 Example record from smartphone sensors

| Time | Accelerometer (x axis) | Gravity sensor (y axis) | Gyroscope (x axis) | Magnetic Sensor (y axis) | … |
|---|---|---|---|---|---|
| 20191220 18:14:37 | -0.7258301 | -0.2037209 | -0.0370554 | -20.8049773 | … |
| 20191220 18:14:37 | -3.4959106 | 9.5317275 | 3.8768861 | -34.7236633 | … |
| 20191220 18:14:38 | 33.5924381 | 9.6923434 | 6.6337223 | -38.7886051 | … |
| … | … | … | … | … | … |
| 20191220 18:14:37 | -11.309281 | 9.6512161 | -4.4452780 | -26.5205389 | … |

**2.2 Feature extraction**

**2.2.1 Temporal feature extraction based on mobile phone sensor data**

In this study, a basic statistical-based feature extraction algorithm (Wang *et al.* 2006, Zhuang *et al.* 2019) was used to calculate the time-series features such as mean, variance, extreme value, entropy, and zero-crossing rate for different dimensional time series data for the six sensors collected. A total of 540-dimensional motion feature vectors were extracted.

In order to accurately reflect the human motion style and reduce the redundancy of feature vectors, this study uses the average precision reduction method to select five of the 540-dimensional motion feature vectors as the final motion feature vectors. The mean accuracy reduction method

(Lin and Jeon 2006) is an assessment method that sets the value of a variable as a random number and keeps other variables constant, and estimates the importance of the variable by analyzing the increase in error of the model after changing the variable through the error propagation formula (Cutler *et al.* 2011).

**2.2.2 Movement pattern extraction based on GPS data**

Studies have demonstrated that coupling multi-sensor data and GPS data for fine-scale human activity recognition has higher accuracy than using a single sensor (e.g. accelerometer) (Zhang and Poslad 2012). In this study, the movement velocity and acceleration of the movers were calculated based on the GPS data as the movement characteristics of the volunteers. Among them, motion velocity can better distinguish whether the volunteers' activities are with the aid of mobile vehicles (e.g., bicycles and electric bicycles) and has been shown to better distinguish human activity patterns (Zhang and Poslad 2014); And acceleration can better reflect whether human movement involves frequent velocity changes and is also commonly used to identify human movement patterns (Guinness 2015). The moving velocity is calculated as shown below.

$$v = \frac{2R * arcsin\sqrt{\left(sin\frac{\Delta A}{2}\right)^2 + cos(lat1\_ra) * cos(lat2\_ra) * \left(sin\frac{\Delta B}{2}\right)^2}}{t_2 - t_1}$$

where $lat1\_ra$ and $lat2\_ra$ are the angular representation of the dimension of the two points, $\Delta A$ is the difference in dimension of the two points, $\Delta B$ is the difference in longitude of the two points, $t_1$ and $t_2$ are the moments between the volunteers to the two points, and R denotes the radius of the Earth.

The moving acceleration is calculated as follows:

$$a = \frac{v_2 - v_1}{t_2 - t_1}$$

where $v_1$ and $v_2$ are the velocities calculated from GPS data at the two nearby points, and $t_1$ and $t_2$ are the moments corresponding to these two points.

**2.3 RunnerDNA**

In molecular biology, a gene is a part of the DNA sequence that determines the expression of a biological trait (Br *et al.* 2002). Since there are significant differences in different individuals' movement styles (Okamoto *et al.* 2014), we can consider the different movement styles of different individuals as distinct expressions of the trait. Human sports styles are determined by mixed factors, which are difficult to quantify in terms of clear indicators (Statham and Hoppenbrouwers 2017, Williams and Ziff 1991). In order to clearly understand the differences in human movement styles, this study uses five motion feature vectors extracted by the mean accuracy reduction method to form five intuitive, interpretable, and easily measurable indicators. These indicators form an individual-scale RunnerDNA to quantitatively describe and measure the differences in movement styles of different users. The five indicators are calculated in Table 3. The value of each indicator is normalized and mapped to 0-5 using min-max normalization. The bigger the indicator values, the more prominent of individual characteristic.

Table 3 Five indicators of RunnerDNA

| | Formula | Parameters | Description |
|---|---|---|---|
| **Balance** | $RMSE_{direction} = \sqrt{\frac{\sum_{i=1}^{n}(zDirection_i - zPrediction_i)^2}{n}}$ | Where, $zDirection_i$ represents the time series data of the z-axis direction sensor, and $zPrediction_i$ represents the value of the time series data fitted with a linear function. | Root Mean Square Error (RMSE) is an effective indicator to describe the distribution of data and is sensitive to outliers(Chai and Draxler 2014). If the RMSE value after linear fitting of the z-axis data of the direction sensor is small, it means that |
| | | | the smartphone shakes less on the thigh plane when the user exercises, and then the user's balance is strong. |
| Stride | $d_{ij} = max\lvert x_{i+k}(t) - x_{j+k}(t)\rvert$<br>$\varphi^m(t) = \dfrac{1}{n-m+1}\displaystyle\sum_{i=1}^{n-m+1} lnC_i^m(t)$<br>$ApEn(t) = \varphi^m(t) - \varphi^{m+1}(t)$ | Using m as the window, the time series is divided into k=n-m+1 series, and $x_t$ is a certain sub-sequence of the time series data of the x-axis direction sensor. $d_{ij}$ is the maximum value of the absolute value of the interpolation of the corresponding elements of the two vectors; $C_i^m(t)$ is the ratio of the number in the table greater than the specified threshold to n-m+1. $\varphi^m$ is the logarithmic average calculated based on k $C_i^m(t)$. $ApEn(t)$ is the approximate entropy of the time series data. | Approximate entropy can be used to express the degree of chaos in a time series(Pincus 1991). If the approximate entropy of the x-axis data of the direction sensor is larger, the time sequence is more complicated, and each frequency in the waveform is more, and then the user's stride is larger. |
| Steer | $RMSE_{linear}$<br>$= \sqrt{\dfrac{\sum_{i=1}^{n}(xLine_i - xPrediction_i)^2}{n}}$ | Where, $xLine_i$ represents the time series data of the x-axis linear acceleration sensor, and $xPrediction_i$ represents the value of the time series data fitted by the cubic fitting function. | If the RMSE value of the linear acceleration sensor's x-axis direction data after three fittings is small, it means that the acceleration changes drastically when the user moves and turns, and then the user has a strong steering ability. |

| | | | | |
|---|---|---|---|---|
| **Stability** | | $L_{linear} = -\log \left( \prod_{i=1}^{n} \frac{1}{\sqrt{2\pi}\sigma} e^{-\frac{(zLiner_i - u)^2}{2\sigma^2}} \right)$ | Where, $zLiner_i$ represents the time series data of the z-axis linear acceleration sensor; $\mu$ and $\sigma$ are the mean and variance of the series data. | The negative log-likelihood function(Williams and Rasmussen 2006) can describe the similarity of data in a process. If the value is large, it means that the sequence similarity is low, and the regularity of the acceleration change is not strong, and the user's stability is poor. |
| **Amplitude** | | $L_{acceleration} = -\log \left( \prod_{i=1}^{n} \frac{1}{\sqrt{2\pi}\sigma} e^{-\frac{(yAcceleration_i - u)^2}{2\sigma^2}} \right)$ | Where, $yAcceleration_i$ represents the time series data of the y-axis linear acceleration sensor; $\mu$ and $\sigma$ are the mean and variance of the series data. | If the value is large, it indicates that the sequence similarity is low, and the y-axis acceleration changes significantly, and then it is known that the user's amplitude is large. |

**2.3 Sport activity recognition and user identification based on RunnerDNA and Random forest**

2.3.1 Random forest models for human activity recognition and user identification

Compared with most machine learning methods, the random forest algorithm not only has higher accuracy but also can be used to evaluate the importance of variables. It can automatically generate unbiased estimates and better balance the errors of unbalanced data sets (Cutler *et al.* 2011, Denisko and Hoffman 2018). Since the human activity style is determined by multiple mixed factors, and the relationship between them is nonlinear, this study uses random forest algorithm to construct a model to recognize sport activity and identify user identity.

Based on the five indicators of RunnerDNA, the random forest algorithm is used to build a user sports activity recognition model and an individual identification model, which are used to classify different sports activities and estimate the data producers corresponding to the different movement data under the sports activity of running.

2.3.2 Accuracy assessment and significance test

To verify the accuracy of the model, this study evaluates the effectiveness of motion pattern recognition and motion identity identification using the accuracy rate (Wan et al. 2020, Zhou et al. 2020). The accuracy rate was measured as follows:

$$Accuracy = \frac{TP + TN}{N}$$

where TP (True Positive) denotes the correctly classified positive class, FP (False Positive) denotes the incorrectly classified negative class, FN (False Negative) denotes the incorrectly classified positive class, and TN (True Negative) denotes the correctly classified negative class. N denotes the total amount of sample data, which is the sum of TP, FN, FP, and TN.

In the process of random forest algorithm to generate each tree, its training set is obtained by Bagging sampling method, only 63% of the data are repeatedly sampled, while the remaining data will not be sampled, which is called out-of-bag (OOB) data (Breiman 1996). As an unbiased estimate of the generalization error of Random forest, OOB estimation can be a good measure of the generalization error of the random forest algorithm(Breiman 2001). Therefore, this study used OOB Error to evaluate the prediction accuracy of models to recognize movement identity and sports activity.

To further verify from a statistical point of view whether the five major indicators of RunnerDNA are significantly different between different sports patterns and users of different genders, this study uses independent sample t-tests for significance analysis. The t value can be measured as follows:

$$t = \frac{\overline{X_1} - \overline{X_2}}{\frac{(n_1-1)S_1^2 + (n_2-1)S_2^2}{n_1+n_2-2}(\frac{1}{n_1}+\frac{1}{n_2})}$$

Where, $\overline{X_1}$ and $\overline{X_2}$ are the mean values of the two samples, $S_1^2$ and $S_2^2$ are the variances of the two samples, and $n_1$ and $n_2$ are the two sample sizes. By calculating the test statistics and querying the corresponding threshold table to determine the P-value, we can test whether there are significant differences between the five indicators of RunnerDNA in different sports activities and different genders.

## 3. Results
### 3.1 Accuracy and significance test

In this study, the motion pattern and motion identity identification models are constructed based on RF. To further evaluate the model effectiveness, we selected KNN, SVM, and NB for comparison experiments, and the obtained accuracies are shown in Table 4. The results show that the random forest model-based models achieve the highest accuracy in recognizing the user's activity pattern and distinguishing the identity of each user in the motion pattern of running.

Table 4 Model accuracy in activity recognition and individual identification

| Accuracy / Algorithm | User sports activity recognition model | Individual identity identification model |
|---|---|---|
| RF | 0.679 | 0.511 |
| KNN | 0.593 | 0.333 |
| SVM | 0.630 | 0.381 |
| NB | 0.494 | 0.429 |

The confusion matrix of the result sports activity recognition is shown in Table 5. As for the activity recognition model, the accuracy of out-of-bag estimation is 0.679, and the overall accuracy of classification is 0.891, with the Kappa coefficient is 0.831. The results show that the random forest model fits well, that predicted values have a high consistency with the target value, which means that the proposed model can effectively distinguish the user's different sports activities. For the identification model, the accuracy of the out-of-bag estimation is 0.511, which means the model can reveal the user's movement style through the sport data, and effectively identify the running user.

Table. Confusion matrix of the pattern identification result
(Columns describe actual labels; rows describe sports activity)

| Sports activity | Biking | E-Bike riding | Walking | Running |
|---|---|---|---|---|
| Biking | 0.625 | 0.018 | 0 | 0.007 |
| E-Bike riding | 0.219 | 0.909 | 0 | 0.022 |
| Walking | 0.062 | 0 | 0.905 | 0.029 |
| Running | 0.094 | 0.073 | 0.095 | 0.942 |

We conducted a t-test analysis of the five indicators, and the results are shown in Table 6. It can be found that walking and running have significant differences in balance, stride, and stability; Biking and walking have significant differences in stride length indicators; Electric bike riding and walking have significant differences in stride and steering indicators; Electric bike riding and running have significant differences in stability indicator. It shows that the five indicators of RunnerDNA of different sports modes have significant differences, and RunnerDNA can better describe different sports activities.

Table 4 Significance analysis of five indicators of RunnerDNA in different sport activity

| Indicator | Sports activity I | Sports activity II | Mean and standard deviation of sports activity I | Mean and standard deviation of sports activity II | t | P |
|---|---|---|---|---|---|---|
| Balance | Biking | E-Bike riding | 2.013, 0.521 | 1.995, 0.515 | 0.159 | 0.874 |
| | Biking | Walking | 2.013, 0.521 | 2.132, 0.715 | -0.781 | 0.437 |
| | Biking | Running | 2.013, 0.521 | 2.759, 0.970 | -4.144 | 5.410 |
| | E-Bike riding | Walking | 1.995, 0.515 | 2.132, 0.715 | -1.091 | 0.278 |
| | E-Bike riding | Running | 1.995, 0.515 | 2.759, 0.970 | -5.492 | 1.262 |
|  | Walking | Running | 2.132, 0.715 | 2.759, 0.970 | -3.881 | 0.0001*** |
| Stride | Biking | E-Bike riding | 2.539, 0.434 | 2.516, 0.448 | 0.230 | 0.819 |
|  | Biking | Walking | 2.539, 0.434 | 2.844, 0.607 | -2.379 | 0.020** |
|  | Biking | Running | 2.539, 0.434 | 3.155, 0.782 | -4.239 | 3.703 |
|  | E-Bike riding | Walking | 2.516, 0.448 | 2.843, 0.607 | -3.041 | 0.003*** |
|  | E-Bike riding | Running | 2.516, 0.448 | 3.155, 0.782 | -5.652 | 5.742 |
|  | Walking | Running | 2.844, 0.607 | 3.155, 0.782 | -2.373 | 0.019** |
| Steer | Biking | E-Bike riding | 1.269, 0.202 | 1.215, 0.169 | 1.304 | 0.196 |
|  | Biking | Walking | 1.269, 0.202 | 1.376, 0.292 | -1.763 | 0.082* |
|  | Biking | Running | 1.269, 0.202 | 2.429, 0.882 | -7.26 | 1.361 |
|  | E-Bike riding | Walking | 1.215, 0.169 | 1.376, 0.292 | -3.389 | 0.001*** |
|  | E-Bike riding | Running | 1.215, 0.169 | 2.429, 0.882 | -10.022 | 3.012 |
|  | Walking | Running | 1.376, 0.292 | 2.429, 0.882 | -7.596 | 1.671 |
| Stability | Biking | E-Bike riding | 2.592, 0.422 | 2.512, 0.080 | 1.368 | 0.175 |
|  | Biking | Walking | 2.592, 0.422 | 2.523, 0.082 | 1.039 | 0.302 |
|  | Biking | Running | 2.592, 0.422 | 2.648, 0.404 | -0.689 | 0.492 |
|  | E-Bike riding | Walking | 2.512, 0.080 | 2.523, 0.082 | -0.684 | 0.495 |
|  | E-Bike riding | Running | 2.512, 0.080 | 2.648, 0.404 | -2.457 | 0.015** |
|  | Walking | Running | 2.523, 0.082 | 2.648, 0.404 | -1.989 | 0.048** |
| Amplitude | Biking | E-Bike riding | 3.373, 0.375 | 3.47, 0.239 | -1.421 | 0.159 |
|  | Biking | Walking | 3.373, 0.375 | 3.475, 0.254 | -1.385 | 0.170 |
|  | Biking | Running | 3.373, 0.375 | 3.400, 0.459 | -0.301 | 0.764 |
|  | E-Bike riding | Walking | 3.467, 0.239 | 3.475, 0.254 | -0.147 | 0.884 |
|  | E-Bike riding | Running | 3.467, 0.239 | 3.400, 0.459 | 1.034 | 0.303 |
|  | Walking | Running | 3.475, 0.254 | 3.400, 0.459 | 1.017 | 0.311 |

"*" p <0.100

"**" p<0.050

"***" p<0.01

## 3.2 The analysis of RunnerDNA

To further mining for the rich information of individual sports style contained in RunnerDNA, this study quantified RunnerDNA of running men and women to explore the gender differences. The t-test results of the five indicators of RunnerDNA by gender are shown in Table 7.

Table 5 The t test of RunnerDNA for runners of different genders

| Indicator | Gender | Mean | SD | t | P |
|---|---|---|---|---|---|
| Balance | Women | 2.013 | 1.167 | 2.316 | 0.022** |
|  | Men | 2.396 | 0.782 |  |  |
| Stride | Women | 2.747 | 0.979 | 1.387 | 0.167 |
| | Men | 2.503 | 1.033 | | |
| Steer | Women | 2.181 | 0.949 | 4.529 | 1.274 |
| | Men | 2.827 | 0.727 | | |
| Stability | Women | 1.519 | 0.155 | -0.529 | 0.597 |
| | Men | 1.489 | 0.408 | | |
| Amplitude | Women | 0.909 | 0.596 | 1.989 | 0.048** |
| | Men | 1.104 | 0.541 | | |

"*" p <0.100

"**" p<0.050

"***" p<0.01

Table 7 shows that among all the volunteers in this study, men and women don't have a significant difference in stride, steer, and stability. At the same time, men are significantly higher than women in both balance and amplitude. Among them, the mean values of balance for men and women are 2.396 and 2.013, respectively, which is consistent with previous research (Sardroodian and Hosseinzadeh 2020); Simth's study on gender differences in the change of the center of mass during human walking found that the vertical displacement was smaller in women than in men (Smith et al. 2002), which coincides with the higher amplitude in men than in women in this study (mean value 1.104 for men and 0.909 for women).

**3.3 Coupling GPS data and RunnerDNA for mobile environment awareness**

To further explore the factors affecting the accuracy of motion pattern recognition and motion identity identification models, this study couples RunnerDNA and the motion features extracted from GPS data to further distinguish human motion patterns and motion identities. In this section, some indoor motion data and inaccurate positioning data were excluded from the original data, and 194 data (23 bicycles, 31 motorcycles, 29 walks, 111 runs) from 17 volunteers were analyzed.

The results show that the sport activity and user identity identification model based on random forest can effectively identify human activity patterns and identities by coupling the motion features extracted from RunnerDNA and GPS data. The accuracy of the sport activity recognition model is 0.830, which is significantly higher than that of the motion pattern recognition model based on RunnerDNA only (0.679). This shows that the movement characteristics can be well extracted and recognized when the human body is moving, so that the movement patterns of the human body can be recognized.

The accuracy of the motion identity identification model constructed by coupling RunnerDNA and the motion features extracted from GPS data is 0.514, which is not significant higher than the model constructed only based on RunnerDNA (0.511). This indicates that the motion features cannot fully reflect the motion preference and motion style of different individuals. Results show that RunnerDNA can fully explore the motion posture and motion style features of different individuals, which is the indicator that can truly reflect the difference of human movement style.

## 4. Discussion

Current research is usually difficult to integrate multi-source sensor data, resulting in challenges of highly accurate activity recognition (Chen and Shen 2017). Meanwhile, due to the lack of multi-sensor data and effective recognition models, previous studies failed to recognize human activity and identity at the individual scale. To address those problems, this study proposed the RunnerDNA, using smartphone multi-sensor data, defined five indicators to describe movement styles of people, which are intuitive, easy- to- measure, and interpretable. Based on five indicators of RunnerDNA, random forest models were developed to sports activity recognition and identification.

By collecting multi-sensor data from thirty- three volunteers in different sports modes, the experiment was conducted based on the concept of RunnerDNA. Four classifiers, RF, KNN, SVM, and NB were selected to compare the accuracy of sport activity recognititon

The accuracy of four classifiers, RF, KNN, SVM and NB, was compared for human sport activity recognition and user identity recognition, and the RF has the best performance for both the movement pattern recognition model and the user identity identification. The results show that the accuracy of the sports activity recognition model reached 0.679, and the accuracy of the movement individual identification model reached 0.511. This indicates that the model can accurately determine the user's sports activity and recognizes running users' identity.

According to the result of significant analysis of five indicators, we found that balance indicator has a significant impact on distinguishing between walking and running; stride indicator has a significant effect on distinguishing between biking and walking, E-Bike riding and walking, walking and running Function; Steering indicator has a significant effect on distinguishing between E-Bike riding and walking; Stability indicator has a significant effect on distinguishing between E-Bike riding and running, and walking and running.

In this study, the morphological characteristics data such as height and weight of each user were collected in detail, and the five indicators of RunnerDNA were used to identify the differences between men and women users' sports styles to differentiate the users' genders. The results show that RunnerDNA can accurately reveal the sports styles of users of different genders. We also conducted significance tests on the five indicators of running volunteers of different genders, and the results showed that men were significantly higher than women in terms of balance indicator

(mean value of 2.396 for men and 2.013 for women) and amplitude indicator (mean value of 1.104 for men and 0.909 for women). This indicates that RunnerDNA can recognize users' movement styles and identities at the individual scale and better identify users' sports activities and explore the differences in movement styles of different genders.

The accuracy of human movement pattern recognition by coupling GPS-based motion features data and RunnerDNA reached 0.830, and the accuracy of user identity identification reached 0.514. This indicates that the GPS data-based movement features can fully exploit the changes of the moving environment of volunteers and have better effect in distinguishing the movement patterns of human body. In terms of movement identity identification, the motion features cannot measure the movement style and preference of volunteers well, while RunnerDNA can distinguish the movement identity of different users in terms of movement posture and style by mining the fine movement style differences of users.

In addition, since the GPS data collection involves the privacy of volunteers to a certain extent, it has certain limitations in practical application and promotion. RunnerDNA can be widely used to explore and analyze the movement style differences of users at individual scale.

The proposed model has several limitations. For example, due to the privacy issue, the amount of data collected is not enough. Thus, this study failed to use big data to mine deeper information from user behavior. Besides, since the volunteers in this study are all college students, the data source does not have enough coverage in terms of age and physical fitness. For example, when using RunnerDNA to analyze the gender difference in sports styles, it is difficult to exclude the influence of height and weight since men and women usually have a large difference in these two characteristics. During data collection, we specified that the volunteers move at their respective normal speeds, so the effect of movement speed on the results could not be completely ignored when exploring differences in amplitude between males and females. Therefore, in future research, we will further improve the quality and quantity of data, increase the coverage of data sources, explore the sports styles of users of all ages and physical fitness, and enhance the model's robustness.

## 5. Conclusion

Using smartphone multi-sensor data, this study proposed RunnerDNA and build RF-based models for sports activity recognition and identification. Based on that, the further analysis of movement style difference between women and men was conducted. The result showed that our model has a good performance in both sports activity recognition and identification, with accuracy is of 0.679 and 0.511, respectively. Also, we found that the balance indicator and acceleration indicator of men is significantly higher than women. The accuracy of human activity recognition coupled with GPS data reached 0.830, and the accuracy of user identity identification reached 0.514, proving the effectiveness of RunnerDNA in recognizing human posture. Our study proposes the RunnerDNA, which is measured by only five indicators, based on the recognition of human activity and identity, the individual's sports style during running is further measured. This research is of great significance for the further intuitive understanding of sports activities and personalized features at the individual scale.

**Funding**

This work was supported by the National Key R&D Program of China (Grant No. 2019YFB2102903) and the National Natural Science Foundation of China (Grant No. 41801306 and 41671408), and China Scholarship Council.

# References

Al Kalaa, M.O. and Refai, H.H., 2015. Selection probability of data channels in Bluetooth Low Energy. In *2015 International Wireless Communications and Mobile Computing Conference (IWCMC)*IEEE), 148-152.

Br, A., Johnson, A. and Lewis, J., 2002. Molecular biology of the cell 4th ed. *New York: Garland Science*.

Breiman, L., 1996. Out-of-bag estimation.

Breiman, L., 2001. Random Forest. *MACHINE LEARNING*, 45, 5-32.

Chai, T. and Draxler, R.R., 2014. Root mean square error (RMSE) or mean absolute error (MAE)?–Arguments against avoiding RMSE in the literature. *Geoscientific Model Development*, 7(3), 1247-1250.

Cutler, A., Cutler, D.R. and Stevens, J.R., 2011. Random forests. *MACHINE LEARNING*, 45(1), 157-176.

Denisko, D. and Hoffman, M.M., 2018. Classification and interaction in random forests. *Proceedings of the National Academy of Sciences*, 115(8), 1690-1692.

Doshi-Velez, F. and Kim, B., 2017. Towards a rigorous science of interpretable machine learning. *arXiv preprint arXiv:1702.08608*.

Dutta, A., et al., 2018. Identifying free-living physical activities using lab-based models with wearable accelerometers. *SENSORS*, 18(11), 3893.

Fugiglando, U., et al., 2017. Characterizing the" driver dna" through can bus data analysis. In *Proceedings of the 2nd ACM International Workshop on Smart, Autonomous, and Connected Vehicular Systems and Services*, 37-41.

González, S., et al., 2015. Features and models for human activity recognition. *NEUROCOMPUTING*, 167, 52-60.

Guinness, R.E., 2015. Beyond where to how: A machine learning approach for sensing mobility contexts using smartphone sensors. *SENSORS*, 15(5), 9962-9985.

Han, Y., et al., 2016. A self-powered insole for human motion recognition. *SENSORS*, 16(9), 1502.

Kaghyan, S. and Sarukhanyan, H., 2013. Accelerometer and GPS sensor combination based system for human activity recognition. In *Ninth International Conference on Computer Science and Information Technologies Revised Selected Papers*IEEE), 1-9.

Kay, M., Santos, J. and Takane, M., 2011. mHealth: New horizons for health through mobile technologies. *World Health Organization*, 64(7), 66-71.

Kwapisz, J.R., Weiss, G.M. and Moore, S.A., 2011. Activity recognition using cell phone accelerometers. *ACM SigKDD Explorations Newsletter*, 12(2), 74-82.

Lin, Y. and Jeon, Y., 2006. Random Forests and Adaptive Nearest Neighbors. *Publications of the American Statistical Association*, 101(474), 578-590.

Martinez-Hernandez, U., Mahmood, I. and Dehghani-Sanij, A.A., 2017. Simultaneous Bayesian recognition of locomotion and gait phases with wearable sensors. *IEEE SENSORS JOURNAL*, 18(3), 1282-1290.

Okamoto, T., et al., 2014. Extraction of person-specific motion style based on a task model and imitation by humanoid robot. In *2014 IEEE/RSJ International Conference on Intelligent Robots and Systems*IEEE), 1347-1354.

Paudyal, P., Banerjee, A. and Gupta, S.K., 2016. Sceptre: a pervasive, non-invasive, and programmable gesture recognition technology. In *Proceedings of the 21st International Conference on Intelligent User Interfaces*, 282-293.

Pincus, S.M., 1991. Approximate entropy as a measure of system complexity. *Proceedings of the National Academy of Sciences*, 88(6), 2297-2301.

Rasekh, A., Chen, C. and Lu, Y., 2014. Human activity recognition using smartphone. *arXiv preprint arXiv:1401.8212*.

Sardroodian, M. and Hosseinzadeh, M., 2020. Gender differences in the spatial–temporal variability between walking and running. *Sport Sciences for Health*, 16(1), 123-127.

Smith, H.J., et al., 2019. Efficient Neural Networks for Real-time Motion Style Transfer. *Proceedings of the ACM on Computer Graphics and Interactive Techniques*, 2(2), 1-17.

Smith, L.K., Lelas, J.L. and Kerrigan, D.C., 2002. Gender differences in pelvic motions and center of mass displacement during walking: stereotypes quantified. *Journal of women's health & gender-based medicine*, 11(5), 453-458.

Statham, A. and Hoppenbrouwers, M.B., 2017. Method and system for feedback on running style field and background of the invention. Google Patents.

Su, X., Tong, H. and Ji, P., 2014. Activity recognition with smartphone sensors. *TSINGHUA SCIENCE AND TECHNOLOGY*, 19(3), 235-249.

Uddin, M.T. and Uddiny, M.A., 2015. Human activity recognition from wearable sensors using extremely randomized trees. In *2015 International Conference on Electrical Engineering and Information Communication Technology (ICEEICT)*IEEE), 1-6.

Wan, S., et al., 2020. Deep learning models for real-time human activity recognition with smartphones. *Mobile Networks and Applications*, 25(2), 743-755.

Wang, X., Smith, K. and Hyndman, R., 2006. Characteristic-based clustering for time series data. *DATA MINING AND KNOWLEDGE DISCOVERY*, 13(3), 335-364.

Williams, C.K. and Rasmussen, C.E., 2006. *Gaussian processes for machine learning*.MIT press Cambridge, MA).

Williams, K.R. and Ziff, J.L., 1991. Changes in distance running mechanics due to systematic variations in running style. *JOURNAL OF APPLIED BIOMECHANICS*, 7(1), 76-90.

Yosinski, J., et al., 2015. Understanding neural networks through deep visualization. *arXiv preprint arXiv:1506.06579*.

Yuan, L., et al., 2017. A highly efficient human activity classification method using mobile data from wearable sensors. *International Journal of Sensor Networks*, 25(2), 86-92.

Zhang, Z. and Poslad, S., 2012. Fine-grained transportation mode recognition using mobile phones and foot force sensors. In *International Conference on Mobile and Ubiquitous Systems: Computing, Networking, and Services*Springer), 103-114.

Zhang, Z. and Poslad, S., 2014. Improved use of foot force sensors and mobile phone GPS for mobility activity recognition. *IEEE SENSORS JOURNAL*, 14(12), 4340-4347.

Zhou, X., et al., 2020. Deep-learning-enhanced human activity recognition for Internet of healthcare things. *IEEE Internet of Things Journal*, 7(7), 6429-6438.

Zhuang, W., et al., 2019. Design of human activity recognition algorithms based on a single wearable IMU sensor. *International Journal of Sensor Networks*, 30(3), 193-206.